\documentclass[12pt]{article}

\usepackage{graphicx}

\renewcommand{\baselinestretch}{1.3}

\newcommand{\iidsim}{\stackrel{i.i.d}{\sim}}
\newcommand{\corr}{\mbox{corr}}
\def\singlespace{\def\baselinestretch{1}\@normalsize}

\def\singlespace{\def\baselinestretch{1}}

\usepackage{amssymb}
\usepackage{amsmath}
\usepackage{bm}
\usepackage{color}
\usepackage{latexsym}

\newtheorem{thm}{Theorem}
\newtheorem{prop}{Proposition}
\newtheorem{lem}{Lemma}
\newtheorem{exam}{Example}

\newcommand{\bA}{{\bf A}}
\newcommand{\bB}{{\bf B}}

\newcommand{\bI}{{\bf I}}
\newcommand{\bP}{{\bf P}}

\newcommand{\bX}{{\bf X}}
\newcommand{\bx}{{\bf x}}
\bmdefine{\bgamma}{\gamma}
\bmdefine{\bzeta}{\zeta}

\newcommand{\RR}{\mathbb{R}}

\newcommand{\RP}{{\rm P}}

\newcommand{\RE}{{\rm E}}
\newcommand{\var}{{\rm Var}}

\newcommand{\argmax}{\operatornamewithlimits{argmax}}
\newcommand{\argmin}{\operatornamewithlimits{argmin}}

\begin{document}


\begin{center}
{\large \bf
Forward variable selection for sparse ultra-high dimensional varying coefficient models
}
\end{center}

\vspace{0.1in}

 \centerline{\bf Ming-Yen Cheng, Toshio Honda, and Jin-Ting Zhang}



\begin{abstract}

Varying coefficient models have numerous applications in a wide scope of scientific areas. While enjoying nice interpretability, they also allow flexibility in modeling dynamic impacts of the covariates. But, in the new era of big data, it is challenging to select the relevant variables when there are a large number of candidates. Recently several work are focused on this important problem based on sparsity assumptions; they are subject to some limitations, however. We introduce an appealing forward variable selection procedure.  It selects important variables sequentially according to a sum of squares criterion, and it employs an EBIC- or BIC-based stopping rule. Clearly it is simple to implement and fast to compute, and it possesses many other desirable properties from both theoretical and numerical viewpoints. We establish rigorous selection consistency results when either EBIC or BIC is used as the stopping criterion, under some mild regularity conditions. Notably, unlike existing methods, an extra screening step is not required to ensure selection consistency. Even if the regularity conditions fail to hold, our procedure is still useful as an effective screening procedure in a less restrictive setup. We carried out simulation and empirical studies to show the efficacy and usefulness of our procedure.

\end{abstract}



\begin{singlespace}
\begin{footnotetext}
{$^{}$Ming-Yen Cheng is Professor, Department of Mathematics, National Taiwan University, Taipei 106, Taiwan (Email: cheng@math.ntu.edu.tw). Toshio Honda is Professor, Graduate School of Economics, Hitotsubashi University, 2-1 Naka, Kunitachi, Tokyo 186-8601, Japan (Email: t.honda@r.hit-u.ac.jp). Jin-Ting Zhang is Associate Professor, Department of Statistics \& Applied Probability, National University of Singapore, 3 Science  Drive  2, Singapore 117546 (Email: stazjt@nus.edu.sg).
This research is partially supported by the Hitotsubashi International Fellow Program and the Mathematics Division, National Center of Theoretical Sciences (Taipei Office). Cheng is supported by the Ministry of Science and Technology grant MOST101-2118-M-002-001-MY3. Honda is supported by the JSPS Grant-in-Aids for Scientific Research (A) 24243031 and (C) 25400197. Zhang is supported by the National University of Singapore  research grant R-155-000-128-112.
}
\end{footnotetext}
\end{singlespace}


\newpage

\section{Introduction}
\label{sec:intro}
We consider variable selection problem for the 
varying coefficient model defined by
\begin{equation}
Y=\sum_{j=0}^p\beta_{0j}(T) X_j + \epsilon,
\label{eqn:e101}
\end{equation}
where $Y$ is a scalar response variable, $X_0\equiv 1$, $X_1,\ldots,X_p$ are the candidate covariates, $\epsilon$ is the random error, and $T\in [0,1]$. The coefficient functions $\beta_{0j}$, $j=0,1,\ldots,p$, 
are assumed to vary smoothly with $T$, and are non-zero for only a subset of the $p$ candidate covariates.
The variable $T$ is an influential variable, such as age or income in econometric studies, and is sometimes called the index variable. The varying coefficient model is a popular and useful semiparametric approach to modeling data that may not obey the restrictive form of traditional parametric models. In particular, while it retains the nice interpretability of the linear models, it allows good flexibility in capturing the dynamic impacts of the relevant covariates on the response $Y$. In addition, in practical applications, some of the true covariates may have simply constant effects while the others have varying effects. Such situations can be easily accommodated by a variant, the  so called semi-varying coefficient model \cite{XZT2004, ZLS2002}. Furthermore, model (\ref{eqn:e101}) has been generalized to modeling various data types including count data, binary response, clustered/longitudinal data, time series, and so on. We refer to \cite{FZ2008} for a comprehensive review and the  extensive literature.

Due to recent rapid developments in technology for data acquisition and storage, nowadays a lot of high-dimensional data sets are collected in various research fields where varying coefficient models find meanings and applications,
such as medicine, marketing and so on. In such situations, the model used to analyze the data is usually sparse, that is,  the number of true covariates is not large even when the dimension is very large. Therefore, under the sparsity condition, some effective variable
selection procedures are necessary in order to carry out meaningful statistical estimation and inference.
In this regard, the penalized variable selection approach emerged as the mainstream in the recent decade. Existing general penalty functions for sparse (ultra-)high-dimensional models include the Lasso  \cite{Tibshirani1996},  group Lasso  \cite{MGB2008, YL2006}, adaptive Lasso \cite{Zou2006}, SCAD \cite{FL2001} and Dantzig selector \cite{CT2007}.

In ultra-high dimensional cases where the dimensionality $p$ is very large,  selection consistency becomes challenging and nearly impossible for existing variable selection methods to achieve, however. Thus, an additional independence screening step is usually necessary before variable selection is carried out. For example, sure independence screening (SIS) methods are introduced by  \cite{FL2008} and \cite{FS2010} for linear models and generalized linear models respectively, and nonparametric independence screening (NIS) is suggested for additive models by \cite{FFS2011}. Under general parametric models, \cite{FXZ2014} suggested using the Lasso at the screening stage before implementing a local linear approximation to the SCAD (or general folded concave) penalty at the second stage.
In all of the above mentioned  variable selection and independence screening methods, some tuning parameter or threshold value is involved which needs to be determined by the user or by some elaborated means.
Under the considered varying coefficient model (\ref{eqn:e101}), there are some existing work on penalized variable selection in several different setups of the dimensionality $p$, using
the Lasso or folded concave penalties such as the SCAD \cite{AGV2012, Lian2012, NP2010,TWZS2012,  WLH2008, WHL2011, XQ2012}.  In ultra-high dimensional cases, for the independence screening purpose, the Lasso is recommended by \cite{WHL2011} and NIS is considered by several authors \cite{CHLP2014, FMD2014, LLW2014, SYZ2014}. Again, all of these methods require selection of some tuning parameter or threshold value.

More recently, an alternative forward variable selection approach receives increasing attention for linear regression. The literature along this line includes the least angle regression (LAR) \cite{EHJT2004}, the forward iterative regression and shrinkage technique (FIRST) \cite{HZG2009}, the forward Lasso adaptive shrinkage (FLASH) \cite{RJ2011}, and the sequential Lasso (SLASSO) \cite{LC2014}. Such methods enjoy desirable theoretical properties,  including selection consistency, and have advantages from numerical aspects. 
Motivated by the above observations, we propose and investigate thoroughly a forward variable selection procedure for the considered varying coefficient model  in
ultra-high dimensional covariate cases, where the dimensionality can be much larger than
the sample size.
The proposed method is constructed in a spirit similar to the SLASSO \cite{LC2014}, which employes Lasso in the forward selection and uses the EBIC \cite{CC2008} as the stopping criterion. However, the selection criterion of our method is based on the reduction in the sum of squared residuals, instead of the Lasso. This is because our preliminary simulation studies suggested that the proposed one performs better than the analogue of the Lasso for the varying coefficient model considered here.

The stopping rule of the proposed forward selection procedure is based on the analogue of the EBIC \cite{CC2008}, or alternatively the BIC, for the varying coefficient model.
The consistency result of the EBIC for model selection in ultra-high dimensional additive models is
established by \cite{Lian2014} when the number of true covariates $p_0$ is bounded. The paper also assumes some
knowledge of the number of true covariates, which may be unrealistic or difficult to obtain in some cases.
On the other hand, without this kind of knowledge, the number of all possible subsets of the candidate variables 
to be considered is too large and there is no guarantee that EBIC-based model selection will perform properly. Therefore, it makes sense to consider a forward selection procedure, which does not require such prior knowledge, and use the EBIC as the stopping criterion.

Suppose we have $n$ i.i.d. observations $\{(\bX_i,T_i,Y_i)\}_{i=1}^{n}$, where $\bX_i=(X_{i0}, X_{i1},\ldots,X_{ip})$, taken from the varying coefficient model (\ref{eqn:e101}):
\begin{equation}
Y_i=\sum_{j=0}^p\beta_{0j}(T_i) X_{ij} + \epsilon_i, \, i=1,\ldots,n.
\label{eqn:e103}
\end{equation}
In our theoretical study, we deal with the ultra-high dimensional case where
\begin{equation}
\log p =O(n^{1-c_p}/L).
\label{eqn:p}
\end{equation}
Here, $c_p$ is a positive constant and $L$ is the dimension
of the B-spline basis used in the estimation of the coefficient functions. We will give more details on the B-spline basis and specify more conditions on $p$ later in Sections \ref{sec:procedure}
and \ref{sec:theorems}; 
especially see Assumptions B(2) and B(3) for
the conditions on $p$.
Throughout this paper, $\#A$ denotes the number of elements of a set $A$,
and $A^c$ is the complement of $A$.  We write $S_0$ for the set of indexes of the true covariates in model (\ref{eqn:e101}), that is, $\beta_{0j} \not\equiv 0$ for $j \in S_0$
and $\beta_{0j} \equiv 0$ for $j \in S_0^c$.
In addition, we write $p_0$ for the number of true covariates, i.e. $p_0\equiv\#S_0$, and consider the case that
\begin{equation}
p_0 = O( (\log n)^{c_S} )
\label{eqn:e105}
\end{equation}
for some positive constant $c_S$. Here, condition (\ref{eqn:e105}) on $p_0$ is imposed for simplicity of presentation;
it can be relaxed at the expense of restricting slightly the order of the dimension $p$ specified in (\ref{eqn:p}).

Under some assumptions we establish the selection consistency of our forward variable selection method when $p$ can be larger than $n$ and $p_0$ can grow slowly with $n$, as specified in (\ref{eqn:p}) and (\ref{eqn:e105}).
Importantly, this means that no independence screening is required before the proposed variable selection procedure. This nice property may be intuitively correct when dealing with sparse parametric models using methods like the SLASSO \cite{LC2014}. But it is not obvious for varying coefficient models; in model (\ref{eqn:e101}) each of the coefficient functions is modeled nonparametrically and involves $L$ parameters in its spline estimation. We exploit desirable properties of B-spline bases
to drive these strong theoretical results.
Note also that our selection consistency results hold when either the EBIC or the BIC is used in the stopping rule.

Interestingly, contradictory to what is suggested for linear models, our simulation results indicate that for the considered varying coefficient model (\ref{eqn:e101}) the BIC outperforms the EBIC when they are used as the stopping criterion in the forward selection procedure. In fact, the EBIC stopping rule tends to stop the forward selection too early and make it miss some important variables. The reason behind this is that the penalty on adding another variable is too large. 
Some adjustments may be helpful in coping with this issue, but fortunately we can circumvent it by using simply the BIC and our simulation results show it works very well. Another problem worth of further study is whether the EBIC is really better in forward selection; it is to account for the large number of possible choices in model selection, but this issue vanishes in forward selection.

As mentioned earlier, there exist some useful procedures for variable
selection in varying coefficient modeling. Nonetheless, the proposed method has many merits compared to them, from both practical and theoretical viewpoints. First, since the important variables are selected sequentially, the final model has good {\it interpretability} in the sense that we can rank the importance of the variables according to the order they are selected. Second, in practice we may have some {\em a priori} knowledge that certain relevant variables should be included in the model. In this case, we always have the {\it flexibility} to start from any subset that contains them. Third, our method employs reasonable sequential selection and stopping rules, and no tuning parameters or threshold parameters are present, meaning that the implementation and the computation are {\it simple and fast}. Fourth, there is a drastic gain in terms of {\it numeric stability} as no inversion of large matrices is necessary, as long as the number of true covariates $p_0$ is not large. By comparison, existing variable selection methods all require independence screening in advance, but the NIS and the group Lasso tend to choose many covariates in order not to miss any true covariates; thus inversion of large matrices is inevitable. 
(Notice that the spline estimation of each of the coefficient functions involves $L$ number of parameters, which has to diverge to infinity with $n$, and we have only one observation for each subject in the present setup.)
Fifth, same as \cite{CHLP2014}, we improve on the order of $p$ as compared with the conditions in \cite{FMD2014}. In other words, the forward procedure can reduce the dimensionality more effectively. Finally, our method requires {\it milder regularity conditions} than the sparse Riesz condition \cite{WHL2011} and the restricted eigenvalue conditions \cite{BRT2009} for the Lasso, which are related to all the candidate covariates (Then, there may be a large set of ``ill-behaved'' covariates with indexes outside of $S_0$, especially when $p$ is very large).

The assumptions we impose in Section \ref{sec:theorems} for the selection consistency of our method may fail to hold in some cases. Nevertheless, in that case we can still
use the proposed procedure for the purpose of independence screening, under a less restrictive
setup specified in Section \ref{screening}. 
Then, we will successfully reduce the number of covariates to a moderate order. This allows us to identify consistently the true covariates  in the next stage, by applying the group SCAD or the adaptive group
Lasso procedure to the variables that pass the screening. See Sections \ref{screening} and \ref{sec:theorems} for the details. Besides, some of the coefficient functions may be constant i.e. $\beta_{0j}\equiv\mbox{const}$ for some $j\in S_0$.
Under such circumstances, we can carry out some group SCAD or adaptive Lasso procedures
to detect both the constant coefficients and the varying coefficients, as suggested in Section 3
of \cite{CHLP2014}.  We refer to \cite{CHLP2014} for such a two-stage approach, i.e. screening and then structure identification, and the theoretical and numerical justifications. Note that, there are indeed some advantages in using the proposed forward procedure as a screening tool. In particular, it tends to remove more irrelevant variables than NIS approaches do, and thus reducing the dimensionality more effectively. See Section \ref{num:compare} for some numerical comparisons.




This paper is organized as follows. In Section \ref{sec:procedure},
we describe the proposed forward variable selection procedure. At each step, it uses the residual sum of squares resulted from spline estimation of an extended marginal model to determine the next candidate feature, and it uses the EBIC or the BIC to decide whether to stop or to include the newly selected feature and continue.
We state the assumptions and theoretical results in
Section \ref{sec:theorems}. Results of simulation and empirical studies
are presented in Section \ref{sec:simulation}. Proofs of all the theoretical
results are given in Section \ref{sec:proofs}.

%

\section{Method}
\label{sec:procedure}

In this section, we describe the proposed forward feature selection procedure.
\medskip
Before that, we introduce some notation.   We write $\| f \|_{L_2}$ and
$\| f \|_\infty$ for the $L_2$ and sup norm of a function $f$ on
$[0,1]$, respectively. When $g$ is a function of some random variable(s),
we define the $L_2$ norm of $g$ by $\| g \|=[\RE \{ g^2 \}]^{1/2}$. For a
$k$-dimensional vector $\bx$, $|\bx|$ stands for the Euclidean norm
and $\bx^T$ is the transpose. We use the same symbol for transpose of
matrices.

Recall $S_0$ is the set of true covariates in the varying coefficient model
(\ref{eqn:e101}). Suppose that we have selected covariates sequentially
and obtain index sets $S_1, \ldots, S_k$ as follows:
\[
S_1 \subset S_2 \subset \cdots \subset S_k \equiv S \subset S_0.
\]
That is, $S_j$ is the index set of the selected covariates upon the completion of the $j$th step, for $j=1,\ldots,k$.
Note that
$S_1$ can be the empty set $\phi$, $\{ 0 \}$ which corresponds to the intercept function, or some non-empty subset of $S_0$ given according
to some {\em a priori} knowledge. Then, at the current $(k+1)$th step, we need to choose another candidate from $S^c$, and then we need to decide whether we should stop or add it to $S$ and go to the next step.
Our forward feature selection criterion is defined in (\ref{eqn:e253}), and we employ
a version of the EBIC, given in (\ref{eqn:e219}), as the stopping rule.  See \cite{CC2008}
for more details about the EBIC.

\subsection{Extended marginal model}\label{marginal}
In this section, we consider spline estimation of the extended marginal model when we add another index to the current index set $S$, which we will make use of in deriving our forward selection criterion. Hereafter we write $S(l)$ for $S\cup \{ l \}$ for any $l \in S^c$.
Temporarily we consider the following extended marginal model for
$S(l), l\in S^c$:
\begin{equation}
Y = \sum_{j \in S(l)}
\overline \beta_{j}(T)X_{j}+ \epsilon_{S(l)}.
\label{eqn:e201}
\end{equation}
Here, the coefficient functions $\overline \beta_j$, $j \in S(l)$,
are defined in terms of minimizing the following mean squared error with respect to $\beta_j,\ j \in S(l)$,
\begin{equation*}
\RE \Big\{ \Big(
Y - \sum_{j \in S(l)} \beta_{j}(T)X_{j} \Big)^2
\Big\}, 
\end{equation*}
where the minimization is over the set of $L_2$ integrable functions on $[0,1]$.
Note that $\| \overline \beta_j\|_{L_2}$ should be larger when $j\in S_0 - S$ than when $j\in S_0^c$.
We will impose some assumptions on these coefficient functions
later in this section and in Section \ref{sec:theorems}.

First, we introduce some more notation related to the B-spline basis
used in estimating the extended marginal model
(\ref{eqn:e201}). 
Let $\bB(t)$ denote the $L$-dimensional equi-spaced B-spline basis on $[0,1]$.
We assume
that $L=c_Ln^{\kappa_L}$ where $\kappa_L \ge 1/5$. The order of the B-spline
basis should be taken larger than or equal to two, under our smoothness assumptions
on the coefficient functions in model (\ref{eqn:e201}). Assumptions B(4)-(5) given in Section \ref{sec:theorems}
ensure that we can approximate the coefficient functions with
the B-spline bases. See \cite{Schumaker2007} for the definition
of B-spline bases. We write
\begin{align*}
\bm{W}_{ij} & = \bB(T_i)X_{ij}\in \RR^L, \quad
\bm{W}_{iS}= (\bm{W}_{ij}^T)_{j\in S}^T\in \RR^{L \#S}, \\
\bm{W}_j  & = (\bm{W}_{1j}, \ldots, \bm{W}_{nj})^T  \quad \mbox{and}\quad
\bm{W}_S  = (\bm{W}_{1S}, \ldots, \bm{W}_{nS})^T.
\end{align*}
Note that $\bm{W}_{ij}$ is a vector of regressors in the spline estimation of $\overline\beta_{j}$ in model (\ref{eqn:e201}), and $\bm{W}_j$ and $\bm{W}_S$ are respectively $n\times L$ and $n\times (L \#S) $ matrices. 
Based on the B-spline basis, we can approximate the varying coefficient model (\ref{eqn:e103}) by the following approximate regression model:
\begin{equation}
Y_i = \sum_{j = 0 }^p
\bgamma_{0j}^T \bm{W}_{ij} + \epsilon_{i}', \, \, i=1,\ldots,n,
\label{eqn:e205}
\end{equation}
where $\bgamma_{0j}\in \RR^L$ and $\bgamma_{0j}^T\bB(t) \approx \beta_{0j}(t)$, $j=0,1,\ldots,p$. Similarly, the spline approximation model when the data come from  the extended marginal model (\ref{eqn:e201})  is given by
\begin{equation}
Y_i = \sum_{j \in S(l) }\overline\bgamma_j^T \bm{W}_{ij}
+ \epsilon_{iS(l)}'
= \overline\bgamma_S^T\bm{W}_{iS}+ \overline\bgamma_l^T \bm{W}_{il}  + \epsilon_{iS(l)}', \, \, i=1,\ldots,n,
\label{eqn:e207}
\end{equation}
where $\overline\bgamma_S^T = (\overline\bgamma_j^T)_{j\in S}$ and
$\overline\bgamma_j$, $j\in S(l)$, are defined by minimizing with respect to $\bgamma_j\in\RR^L$, $j\in S(l)$, the following mean squared spline approximation error:
\begin{equation*}
\RE \Big\{
\sum_{i=1}^n \big( Y_i - \sum_{j \in S (l)}\bgamma_j^T \bm{W}_{ij}
\big)^2 \Big\} =
\RE \Big\{ \big|\bm{Y}-\bm{W}_S\bgamma_S -
\bm{W}_l\bgamma_l\big|^2 \Big\}
\end{equation*}
with $\bgamma_S^T = (\bgamma_j^T)_{j\in S}$. Note that
$\overline\bgamma_j^T \bB(t) $ should be close to the coefficient function
$\overline \beta_j(t)$ in the extended marginal model (\ref{eqn:e201}). In particular, when
$l\in S_0$, $\| \overline \beta_l \|_{L_2}$ should
be large enough, and thus $|\overline \bgamma_l |$ should be
also large enough.

We can estimate the vector parameters $\overline\bgamma_j$, $j\in S(l)$, in model (\ref{eqn:e207}) by the ordinary least squares estimates, denoted by $\widehat\bgamma_j$, $j\in S(l)$.
Let $\bm{\widehat W}_{lS}$ and $\bm{\widehat Y}_{S}$
denote respectively the orthogonal projections of $\bm{W}_{lS}$
and $\bm{Y}= (Y_1, \ldots, Y_n)^T$ onto the linear space spanned by the columns of $\bm{W}_S$, that is,
\begin{align*}
\bm{\widehat W}_{lS} & = \bm{W}_S (\bm{W}_S^T
\bm{W}_S )^{-1} \bm{W}_S^T\bm{W}_l
\quad \mbox{and}\quad \bm{\widehat Y}_{S}  = \bm{W}_S (\bm{W}_S^T
\bm{W}_S )^{-1} \bm{W}_S^T\bm{Y}\,.
\end{align*}
Note that  $\bm{\widehat W}_{jS}$ is an $n\times L$ matrix.
Then the ordinary least square estimate  of $\overline \bgamma_l$, denoted by $\widehat \bgamma_l$, can be expressed as
\begin{equation}
\widehat \bgamma_l = ( \bm{\widetilde W}_{lS}^T
\bm{\widetilde W}_{lS})^{-1}
\bm{\widetilde W}_{lS}^T \bm{\widetilde Y}_S,
\label{eqn:e217}
\end{equation}
where
$\bm{\widetilde W}_{jS}  = \bm{W}_j - \bm{\widehat W}_{jS}$ and $\bm{\widetilde Y}_{S} = \bm{Y} - \bm{\widehat Y}_{S}$.
Note that $\widehat \bgamma_l^T \bB(t)$ is the spline estimate of the coefficient function $\overline \beta_l(t)$ in the extended marginal model (\ref{eqn:e201}).

\subsection{Forward feature selection procedure}\label{algorithm}

Recall that at the current step we are given $S$, the  index set of the covariates already selected, and the job is to choose from $S^c$ another candidate and then decide whether we should add it to $S$ or we should not and stop. For the purpose of forward feature selection, we consider the reduction in the sum of squared residuals, or equivalently the difference in the variance estimation, when adding  $l$ to $S$. Specifically, we compute
$\widehat \sigma_S^2 -   \widehat \sigma_{S(l)}^2$, where $\widehat \sigma_Q^2$ is the variance estimate for a subset of covariates indexed by $Q$ given as
\begin{equation}\label{eqn:sig}
\widehat \sigma_Q^2 =\frac{1}{n}
\Big\{
\bm{Y}^T \bm{Y} - \bm{Y}^T \bm{W}_Q ( \bm{W}_Q^T \bm{W}_Q )^{-1}
\bm{W}_Q^T \bm{Y}
\Big\}.
\end{equation}
Using (\ref{eqn:e217}), we can rewrite $\widehat \sigma_S^2 -   \widehat \sigma_{S(l)}^2$ as
\begin{eqnarray}
\widehat \sigma_S^2 -   \widehat \sigma_{S(l)}^2
& = & \frac{1}{n}\big( \bm{\widetilde W}_{lS}^T \bm{\widetilde Y}_{S}\big)^T
\big( \bm{\widetilde W}_{lS}^T \bm{\widetilde W}_{lS} \big)^{-1}
\big( \bm{\widetilde W}_{lS}^T \bm{\widetilde Y}_{S}\big)\nonumber\\
& = &
\widehat \bgamma_l^T \Big( \frac{1}{n}
 \bm{\widetilde W}_{lS}^T \bm{\widetilde W}_{lS} \Big)
\widehat \bgamma_l \, \approx \,
\RE{ \Big\{\big( \overline \beta_l (T)\widetilde X_{lS}\big)^2 \Big\}},
\label{eqn:e251}
\end{eqnarray}
where $\widetilde X_{lS}=X_l - \widehat X_{lS}$
and $\widehat X_{lS}$ is the projection of $X_l$ to
$\big\{ \sum_{j \in S}\beta_j(T)X_j \big\}$ with respect to the $L_2$ norm $\| \cdot \|_{L_2}$.

As noted earlier, if $l \in S_0$ then $\| \overline \beta_l\|_{L_2}$ will be large enough.
Furthermore, $ n^{-1}\bm{\widetilde W}_{lS}^T \bm{\widetilde W}_{lS}$
will have desirable properties under Assumption X(2) given in Section
\ref{sec:theorems}; see Lemma \ref{lem:lem1} for the details. Hence, following from expression (\ref{eqn:e251}) and recalling that $\widehat \bgamma_l^T \bB(t)$
is the spline estimate of $\overline \beta_l(t)$, we choose the candidate index as
\begin{equation}
l^*= \argmin_{l\in S^c} \widehat \sigma_{S(l)}^2 \,.
\label{eqn:e253}
\end{equation}
Then, we have high confidence that $l^*$ belongs to $S_0-S$ provided that the latter is non-empty, and we take $X_{l^*}$ as the next candidate feature.
At first, instead of (\ref{eqn:e253}), we considered choosing
\begin{equation}
l^{\dag}= \argmax_{j\in S^c} \big|  \bm{\widetilde W}_{lS}^T \bm{\widetilde Y}_S\big|
\label{eqn:e215}
\end{equation}
as the next candidate index, as motivated by the sequential Lasso for linear models
proposed by \cite{LC2014}.
However, after some simulation studies we found that, contrary to the nice properties of its counterpart in linear models, (\ref{eqn:e253}) performs better for the varying coefficient model we study.

To determine whether or not to include the candidate feature $X_{l^*}$ in the set of selected ones, we employ the EBIC criterion. Specifically, we define the EBIC of a subset of covariates indexed by $Q$ as the following:
\begin{equation}
\mbox{EBIC}(Q) = n \log (\widehat \sigma_Q^2 ) + \# Q \times L(\log n + 2\eta \log p),
\label{eqn:e219}
\end{equation}
where $\eta$ is a fixed constant and $\widehat \sigma_Q^2$ is given in (\ref{eqn:sig}). Then, at the
current $(k+1)$th step, we should select the new covariate $X_{l^*}$ with $l^*$ defined in (\ref{eqn:e253}), provided that the EBIC decreases when we add $l^*$ to $S$ and form $S(l^*)$. Otherwise, if the EBIC increases, we should not select any more covariates and stop at
the $k$th step.
Note that the EBIC defined in (\ref{eqn:e219}) reduces to the BIC when $\eta$ is taken as 0. And, the theoretical results given in Section \ref{sec:theorems}, in particular the consistency results given in Theorem \ref{thm:thm2}, hold when either the EBIC or the BIC is used as the stopping criterion in the proposed method. In the following, we define formally the proposed forward feature selection algorithm.\\

\noindent{\bf Forward feature selection algorithm.}
\begin{description}
\item{\it Initial step:} Specify $S_1$, which can be taken as the empty set $\phi$, $\{ 0 \}$, or some non-empty subset of $S_0$ chosen based on some {\em a priori} knowledge, and compute $\mbox{EBIC}(S_1)$.
\item{\it Sequential selection:} At the $(k+1)$th step, compute $\widehat\sigma^{2}_{S_k(l)}$ for every $l\in S_k^c$, and find
\[
l^*_{k+1}=\argmin_{l\in S_k^c}\widehat\sigma^2_{S_k(l)}\,.
\]
Then, let $S_{k+1}=S_k\cup \{ l^*_{k+1} \}$ and compute $\mbox{EBIC}(S_{k+1})$. Stop and declare $S_{k}$ as the set of selected covariate indexes if $\mbox{EBIC}(S_{k+1})>\mbox{EBIC}(S_{k})$; otherwise, change $k$ to $k+1$ and continue to search for the next candidate feature.
\end{description}

The forward procedure with the EBIC stopping rule tends to stop a little too early and miss some relevant variables, and we need some kind of
modification when we implement it.
For example, some adjustment of the degrees of freedom will be helpful.
All the details are given in Section \ref{sec:simulation}.

\subsection{Sparsity assumptions}\label{sparisty}

We need some assumptions to establish consistency of
the proposed procedure, especially Assumption B(1) given below.
When conditions B(1)-(2) are not fulfilled, another setup in which we can use the proposed method as a screening approach is given in Section \ref{screening}.
In this paper, $C_1$, $C_2$, $\ldots$ are generic positive constants
and their values may change from line to line. Recall that $S_0$ is the index set of the true variables in model (\ref{eqn:e101}).

\medskip
\noindent
{\bf Assumption B(1)-(2)}
\begin{description}
\item B(1) For some large positive constant $C_{B1}$,
\[
\max_{j \in S_0 - S} \| \overline \beta_j \|_{L_2}
/ \max_{j \in S_0^c} \| \overline \beta_j \|_{L_2}> C_{B1}
\]
uniformly in $S \subsetneq S_0 $. Note that $C_{B1}$ should
depend on the other assumptions on the covariates, specifically
Assumptions {\bf X} and {\bf T} given in Section \ref{sec:theorems}.

\item B(2) Set $\displaystyle{\kappa_n= \inf_{S \subsetneq S_0} \max_{j\in S_0-S}
\|\overline \beta_j \|_{L_2}}$. We assume
\[
\frac{n\kappa_n^2}{L\max\{\log p, \log n\}} > n^{c_\beta}
\quad \text{and} \quad
\kappa_n > \frac{L}{n^{1-c_\beta}}
\]
for some small positive constant $c_\beta$. In addition, if $\eta=0$ in (\ref{eqn:e219}) i.e. if BIC is used, we require that
\[
\frac{L\log n}{\log p} \rightarrow \infty \,.
\]
\end{description}

\medskip
An assumption similar to Assumption B(1) is imposed in \cite{LC2014}
and such assumptions are inevitable in establishing
the selection consistency of forward procedures. These assumptions
ensure that the chosen index $l^*$, given in (\ref{eqn:e253}), will be from $S_0-S$. When such assumptions
fail to hold, our method may choose some covariates from $S_0^c$. However,
these covariates will be removed at the second stage mentioned in
the Introduction. See Section \ref{screening} for more details. 
The first condition in Assumption B(2) is related to the
convergence rate of $\widehat \bgamma_l$, and it ensures that the signals
are large enough to be detected.
If $C_1< \kappa_n <C_2$ for some positive constants $C_1$ and $C_2$, this condition is simply
$\log p < n^{1-c_\beta}/L$ for some small positive constant $c_\beta$, which is fulfilled by assumption (\ref{eqn:p}) on $p$.
A few more assumptions on the coefficient functions $\overline \beta_j(t)$ will be given in
Section \ref{sec:theorems}.
The last condition in Assumption B(2) is to ensure that, when the BIC is used as the stopping criterion, our method can deal with ultra-high dimensional cases. For example, if $L$ is taken of  the optimal order $n^{1/5}$ then $p$ can be taken as $p=\exp(n^{c})$ for any $0<c\leq 1/5$.

\subsection{Forward feature screening}\label{screening}

Some of the assumptions we impose in Section \ref{sec:theorems} may not hold. For example, Assumption B(1) may not hold if some of the irrelevant variables have strong correlation with the true covariates indexed by $S_0$. Thus, such assumptions may be too restrictive in practice, in particular when $p$ is very large and $p_0$ is much smaller than $p$ as specified in (\ref{eqn:p}) and (\ref{eqn:e105}).
In that case, the proposed forward selection procedure may be still used as a forward screening method under certain less restrictive conditions. Then, although some unimportant variables may pass the forward screening, we can utilize some variable selection method to remove them at the next stage.   In this section we discuss the details.

Suppose there is a subset of indexes, denoted by $\overline S_0$, that
contains $S_0$, and the covariates in $\overline S_0$
do not have much correlation
with those in $\overline S_0^c$. To be clear, we specify the conditions
as follows:
\begin{description}
\item[(a)] $S_0 \subset \overline S_0$ and $\# \overline S_0 \le C
\# S_0$ for some positive constant $C$.
\item[(b)] $\displaystyle{\max_{j \in \overline S_0 - S} \| \overline \beta_j \|_{L_2}
\big/ \max_{j \in \overline S_0^c} \| \overline \beta_j \|_{L_2}
\to \infty}$ uniformly for $ S $ satisfying $ S \subsetneq \overline S_0$
and $ S_0 \not\subset S$.
\item[(c)]  Assumption B(2) holds with $ \kappa_n $ replaced with $ \kappa_n'$,
where $ \kappa_n' $ is defined by
\[
\kappa_n'= \inf_{ S } \max_{ j \in S_0-S }
\|\overline \beta_j \|_{L_2},
\]
with $S$ satisfying the same conditions as in (b).
\end{description}

If we replace conditions B(1) and B(2) with conditions (b) and (c), respectively, and if condition (a) holds, then
our procedure given in Section \ref{algorithm} can be used as a forward independence
screening procedure with an effective stopping rule. That is, it will effectively
select all the true covariates indexed by $S_0$, possibly along with some irrelevant ones from those indexed by
$ \overline S_0 - S_0$. See Proposition \ref{prop:prop1} 
given in Section \ref{sec:theorems} for the theoretical justifications.
Those remaining irrelevant covariates will be removed when we apply at the second stage the group SCAD or adaptive group Lasso \cite{CHLP2014,FXZ2014}.

\section{Assumptions and theoretical properties}
\label{sec:theorems}
In this section, we describe technical assumptions, and we present
desirable theoretical properties of the proposed forward procedure
in Theorems \ref{thm:thm1} and \ref{thm:thm2}. Note that
we treat the EBIC and the BIC ($\eta=0$) in a unified way.
The proofs are given in Section \ref{sec:proofs}.

First we describe assumptions on the index variable
$T$ in the varying coefficient model (\ref{eqn:e101}). The following assumption is a standard one
when we employ spline estimation.

\medskip
\noindent
{\bf Assumption T.}\, The index variable $T$ has density function $f_T(t)$ such
that $C_{T1} < f_T(t) < C_{T2}$ uniformly in $t\in[0,1]$, for some positive constants
$C_{T1}$ and $C_{T2}$.

\medskip
We define some more notation before we state our assumptions on the
covariates. Let $\bm{X}_S$ consist of
$\{ X_j \}_{j\in S}$ and then $\bm{X}_S$ is a $\# S $-dimensional
random vector. Note that $\bm{X}_{S(l)}$ is a $(\# S +1) $-dimensional
random vector. For a symmetric matrix $\bA$, we denote the maximum and minimum
eigenvalues respectively by $\lambda_{\rm max}(\bA)$
and $\lambda_{\rm min}(\bA)$, and we define $|\bA|$ as
\[ |\bA| = \sup_{|\bx|=1}|\bA\bx|=
\max\{ |\lambda_{\rm max}(\bA)|, \, |\lambda_{\rm min}(\bA)|\}.
\]

\medskip
\noindent
{\bf Assumption X.}
\begin{description}
\item X(1) There is a positive constant $C_{X1}$ such that
$|X_j| \le C_{X1}$, $j=1,\ldots,p$.
\item X(2) Uniformly in $ S \subsetneq S_0$ and $l\in S^c $,
\[
C_{X2} \le \lambda_{\rm min} ( \RE \{ \bm{X}_{S(l)}\bm{X}_{S(l)}^T | T\})
\le \lambda_{\rm max} ( \RE \{ \bm{X}_{S(l)}
\bm{X}_{S(l)}^T | T\}) \le C_{X3}
\]
for some positive constants $C_{X2}$ and $C_{X3}$.
\end{description}

We use the second assumption X(2) when we evaluate eigenvalues of the matrix
$\RE \{ n^{-1}\bm{W}_{S(l)}^T \bm{W}_{S(l)}\}$.
We can relax Assumption X(1) slightly by replacing $C_{X1}$ with
$C_{X1}(\log n )^{c_X}$ for some positive constant $c_X$. These are standard assumptions in the
variable selection literature.

Assumption {\bf E} below is about the error term $\epsilon$ in our
varying coefficient model (\ref{eqn:e101}). The second condition
E(2) requires that $\epsilon$ should have the sub-Gaussian
property. 
We use it
when we prove the latter half of Theorem \ref{thm:thm2}. This is a standard
assumption in the Lasso literature, for example,
see \cite{BRT2009} and \cite{WHL2011}.

\medskip
\noindent
{\bf Assumption E.}
\begin{description}
\item
E(1) There are positive constants $C_{E1}$ and $C_{E2}$ such
that
\[
\RE \{ \exp (C_{E1}|\epsilon|) | X_1, \ldots, X_p, T \}
\le C_{E2}.
\]
\item
E(2) There is a positive constant $C_{E3}$ such that
$ \RE \{ \exp( u \epsilon ) | X_1, \ldots, X_p, T \}
\le \exp ( C_{E3}u^2/ 2)$ for any $u\in\RR$.
\end{description}

\medskip
We need some additional assumptions on the coefficient functions
$\overline \beta_j$ in the extended marginal model (\ref{eqn:e201}) in order to approximate them
by the B-spline basis. Note that, in Assumptions B(4)-(5) below,
$\overline \beta_j\equiv\beta_{0j}$ for all $j\in S_0$ and
$\overline \beta_j\equiv 0 $ for all $j\in S_0^c$
when $S=S_0$.

\medskip
\noindent
{\bf Assumption B(3)-(5)}.
\begin{description}
\item
B(3) $\kappa_n L^2 \to \infty$ and $\kappa_n = O(1)$,
where $\kappa_n$ is defined in Assumption B(2).

\item
B(4) $\overline \beta_j$ is twice continuously differentiable
for any $j \in S(l)$ for $S \subset S_0$ and $l\in S^c$.

\item
B(5) There are positive constants $C_{B2}$ and $C_{B3}$ such
that $\displaystyle{\sum_{j\in S(l)}\| \overline \beta_j \|_\infty <C_{B2}}$
and $\displaystyle{\sum_{j\in S(l)}\| \overline \beta_j'' \|_\infty <C_{B3}}$
uniformly in $S \subset S_0$ and $l\in S^c$.
\end{description}
\medskip

Theorem \ref{thm:thm1} given below suggests that the forward selection procedure using criterion (\ref{eqn:e253}) can pick up
all the relevant covariates in the varying coefficient model (\ref{eqn:e101}) when $C_{B1}$ in
Assumption B(1) is large enough.

\begin{thm}
\label{thm:thm1}
Assume that Assumptions {\bf T}, {\bf X}, B(1)-(5), and E(1) hold,
and define $l^*$ as in (\ref{eqn:e253}) for any $S \subsetneq S_0$.
Then, with probability tending to 1, there is a positive constant $C_L$ such that
\[
\frac{\big\| \overline\beta_{l^*}\big\|_{L_2}}{\displaystyle{\max_{j \in S_0 - S}}
\big\| \overline\beta_{j}\big\|_{L_2}} > C_L
\]
uniformly in $S$,
and thus we have $l^*\in S_0-S$ for any $S \subsetneq S_0$ when $C_{B1}$ in Assumption B(1)
is larger than $1/C_L$.
\end{thm}

\medskip


Theorem \ref{thm:thm2} given next implies that the proposed forward procedure
will not stop until all of the relevant variables indexed by $S_0$ have been selected, and it does stop when all the true covariates in model (\ref{eqn:e101})
have been selected. Note that in the second result, we have to replace Assumption E(1) with
E(2) in order to evaluate a quadratic form of error terms in the proof.

\begin{thm}
\label{thm:thm2}
Assume that Assumptions {\bf T}, {\bf X}, B(1)-(5), and E(1) hold. Then we have the following results.
\begin{description}
\item
(i) For $l^*$ as in Theorem \ref{thm:thm1}, we have $$\mbox{EBIC}(S(l^*)) < \mbox{EBIC} (S)$$
uniformly in $S \subsetneq S_0 $,  with probability tending to 1.

\item
(ii) If we replace Assumption E(1) with Assumption E(2), then we have
\[
 \mbox{EBIC}(S_0(l)) > \mbox{EBIC} (S_0)
\]
uniformly in $l\in S_0^c$, with probability tending to 1.
\end{description}
\end{thm}

The forward method may also choose some irrelevant covariates if Assumption B(1) fails to hold.
In that case, Proposition \ref{prop:prop1} provides some theoretical results
in the setup described in Section \ref{screening}. Note that some conformable changes
to Assumptions B(3)-(5) and X(2) and the proofs are needed.
See Section \ref{sec:proofs} for the changes in
the proofs.

\begin{prop}\label{prop:prop1}
Consider the setup given in Section \ref{screening}. Under the same conditions in Theorem \ref{thm:thm1} (or Theorem \ref{thm:thm2}), with conformable changes to Assumptions B(3)-(5) and X(2), we have the following results.
\begin{description}
\item (i) The selected index $l^*$ comes only from $\overline S_0$ with
probability tending to 1, as in Theorem \ref{thm:thm1}.
\item (ii)
With probability tending to 1, the proposed forward selection procedure continues the feature selection until all the covariates indexed by $S_0$ are
selected, and it stops the selection when all the covariates indexed by $S_0$
have been selected.
\end{description}
\end{prop}

Proposition \ref{prop:prop1} implies that the proposed forward selection procedure can be used as a forward screening method with an effective
stopping rule. Note that, in this setup, we may select some irrelevant covariates from those indexed by $\overline S_0-S_0$. However, the number of potential covariates will be sufficiently reduced after the forward screening stage.
Thus, we will be able to remove those remaining irrelevant covariates at the next stage, by using
the group SCAD or the adaptive group Lasso \cite{CHLP2014,FXZ2014}.

\section{Simulation and empirical studies}
\label{sec:simulation}

We carried out two simulation studies and a real data analysis based on the well-known Boston housing data to assess the performance of the proposed forward feature selection method with BIC or EBIC as the stopping criterion. For simplicity, we denote these two variants by fBIC and fEBIC respectively.  At the initial step of the forward selection, we let $S_1=\{0\}$ i.e. we start with the model with only the intercept function.
Note that it may happen that the BIC/EBIC drops in one iteration, then increases in the next iteration, and then drops  again. To avoid interference caused by such small fluctuations, we continued the fBIC/fEBIC forward selection until the BIC/EBIC continuously increases for five consecutive iterations. The value of the parameter $\eta$ in the definition  (\ref{eqn:e219}) of EBIC was taken as $\eta=1-\log n/(3\log p)$,  as suggested by \cite{CC2008}. Since the EBIC uses a much larger penalty than the BIC does, it is expected that the fEBIC will select a smaller model than that selected by the fBIC. We could modify the penalty term by adjusting the degrees of freedom or change the value of $\eta$ to a smaller one, but it becomes complicated.

 In the simulation studies, we generated data from the two varying coefficient models
studied by \cite{FMD2014}.  Following the paper, we used the cubic B-spline with  $L = 7$, we set the sample size and the number of covariates as $n = 400$ and $p = 1000$ respectively, and we repeated each of the simulation configuration for  $N = 200$ times.

\subsection{Comparison of fBIC and fEBIC}

In this section,  we compare the finite sample performance of the fBIC and the fEBIC using the two varying coefficient models
studied by \cite{FMD2014}.

\begin{exam}\label{exp1}  Following Example 3 of \cite{FMD2014}, we generated $N$ samples from  the following varying coefficient model:
\[
Y = 2\cdot X_1 + 3\,T\cdot X_2 +  (T+1)^2\cdot X_3 +\frac{4\sin(2\pi T)}{2-\sin(2\pi T)}\cdot X_4+\epsilon,
\]
where $X_j=(Z_j+t_1U_1)/(1+t_1), j=1,2,\cdots, p$, and $T=(U_2+t_2 U_1)/(1+t_2)$, with
$Z_1, Z_2, \cdots, Z_p\iidsim N(0,1), U_1, U_2\iidsim U(0,1)$, and $\epsilon\sim N(0,1)$ being all mutually independent with each other.
\end{exam}

\begin{table}
\footnotesize
\caption{Correlations between the covariates  $X_j$'s and the index variable $T$.}\label{corr.tab}
\begin{center}
\begin{tabular}{|c|c|c|c|c|c|c|} \hline
$[t_1,t_2]$ & $[0,0]$  & $[2,0]$ &$[3,0]$ & $[2,1]$ &$[3,1]$&$[3,2]$\\ \hline
$\corr(X_j,X_k)$ &0&0.25&0.43&0.25 &0.43 &0.43\\ \hline
$\corr(X_j,W)$   &0&0 &0 &0.36 &0.46 &0.59\\ \hline
\end{tabular}
\end{center}
\end{table}

In this example, the number of true covariates $p_0$ is four. The tuning parameters
$t_1$ and $t_2$ are used to control the correlations between the covariates $X_j, j=1,2,\cdots, p$ and the index covariate $T$. It is easy to show that $\corr(X_j,X_k)=t_1^2/(12+t_1^2)$  for any $j\neq k$, and $\corr(X_j,T)=t_1t_2/[(12+t_1^2)(1+t_2^2)]^{1/2}$ independent of $j$.
 Table~\ref{corr.tab} lists the values of the tuning parameters $[t_1,t_2]$ which define six cases of the correlations between the covariates $X_j$'s and the index covariate $T$.  The first case is associated with the situation   when  the $X_j$'s are uncorrelated while they are uncorrelated with $T$.  The second and third cases are associated with those situations when the $X_j$'s are increasingly correlated but they are uncorrelated with $T$. The last three cases are associated with those situations when the $X_j$'s are increasingly correlated and the
 correlations between the $X_j$'s and $T$ are also increasing. These six cases allow us to compare the performance of the fBIC and fEBIC procedures effectively, In the next section, we will also use them to compare the performance of the fBIC with those procedures proposed and studied by \cite{FMD2014}.

\begin{figure}[ptbh]
\centerline{\includegraphics[scale=0.50]{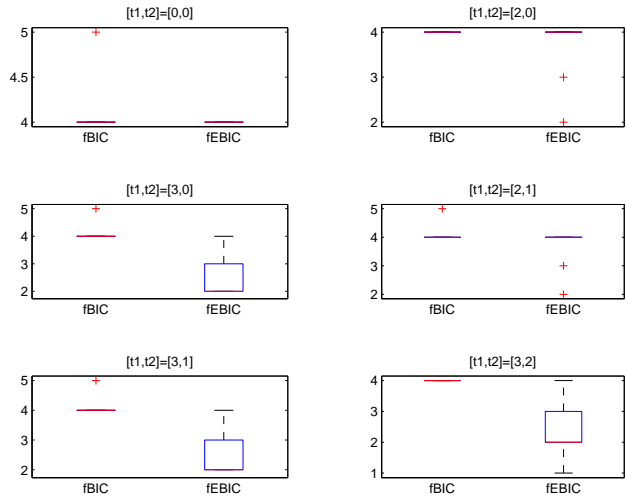}}
\caption{\em Boxplots of the model sizes selected by the fBIC and fEBIC for the varying coefficient model in Example~\ref{exp1}} \label{BICvsEBIC1}
\end{figure}

Figure~\ref{BICvsEBIC1} depicts the boxplots of the model sizes selected by the fBIC and the fEBIC in the six correlation cases. It is seen that in all the six cases,
the fBIC performs very well in terms of  correctly selecting  the right model except that  it occasionally  selects a model with one extra covariate out of the 200 runs. However, generally speaking the fEBIC selects a smaller model as compared to the true model, and it selects all of the four true covariates most of the time only when the correlations between the $X_j$'s and  $T$ are relatively small. As the correlations between the $X_j$'s or the correlations between the $X_j$'s and $T$ increase, the performance of fEBIC becomes worse and it selects a much  smaller model than the correct one most of the time.

The varying coefficient model in Example~\ref{exp1} has only four true  underlying covariates. In the varying coefficient model defined in the following example, there are eight true underlying  covariates.

\begin{exam}\label{exp2}  Following Example 4 of \cite{FMD2014}, we generated $N$ samples from  the following varying coefficient model:
\begin{align*}
Y &=& 3 \,T\cdot X_1 + (T + 1)^2\cdot  X_2 + (T-2)^3\cdot X_3+3(\sin(2\pi T))\cdot X_4\\
  &&  +\exp(T)\cdot X_5+2\cdot X_6+2\cdot X_7+3\sqrt{T}\cdot X_8+\epsilon,
\end{align*}
while $T,\,\bX,\, Y $ and $\epsilon$ were generated in the same way as described in Example~\ref{exp1}.
\end{exam}

\begin{figure}[ptbh]
\centerline{\includegraphics[scale=0.50]{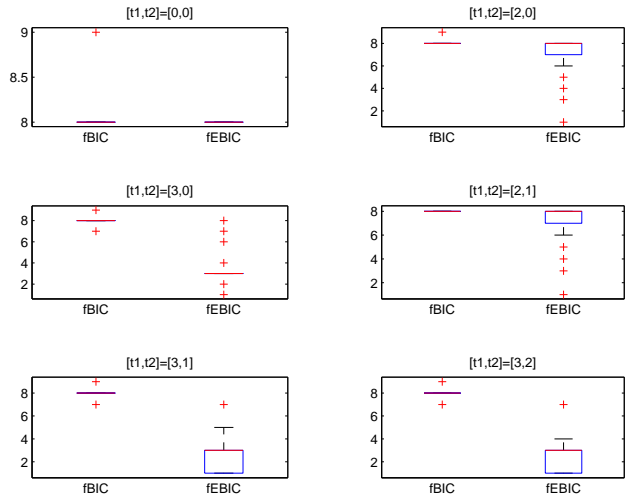}}
\caption{\em Boxplots of the model sizes selected by the fBIC and fEBIC for the varying coefficient model in Example~\ref{exp2}} \label{BICvsEBIC2}
\end{figure}

Figure~\ref{BICvsEBIC2}  shows the boxplots of the model sizes selected by fBIC and fEBIC in the six correlation cases given in Table \ref{corr.tab}, when the data came from the varying coefficient model defined in Example~\ref{exp2}. Again, we observe that in all these six cases, the fBIC performs very well in terms of  correctly selecting  the right model except that  it occasionally  selects a model with one extra or one less  covariate out of the 200 runs. However, the fEBIC  selects a smaller model in general, and it  selects the right model only when the correlations between $X_j$'s and  $T$ are relatively small. Similar to Example 1, when the correlations between $X_j$'s or in the correlations between $X_j$'s and $T$ increase, the performance of fEBIC becomes worse and it selects  a much  smaller model than the right model most of the time.

From the above two examples, we see that the fBIC consistently outperforms the fEBIC substantially. It appears that when a forward selection procedure is used in the considered context, the BIC-based stopping rule is better than the one using EBIC, since the EBIC penalizes the introduction of a new covariate too much and as a result it stops too early. This may seem to contradict with the rational behind the original EBIC designed for linear models. But, for  varying coefficient models the degrees of freedom in the definition of EBIC increases much faster when more variables are introduced to the model. Note also the original EBIC is introduced for model selection, not forward selection. Following the observation that fBIC performs very well numerically and the fact that $\eta$ disappears from it, we prefer the fBIC to the fEBIC for the studied problem.

\subsection{Comparison with the approaches of  Fan, Ma, and Dai (2014)}\label{num:compare}

In this section,  we compare the performance of the fBIC with  that of the  conditional-INIS and the greedy-INIS approaches introduced by \cite{FMD2014}. We consider exactly the
same simulation setups as their Examples 3 and 4 and adopt their simulation results.  Following \cite{FMD2014}, we report the average numbers of true positive (TP) and false positive
(FP) selections, the prediction error (PE), and their robust standard deviations for all the three procedures under consideration, where the prediction error is the mean
squared error calculated on a test dataset of size $n/2 = 200$ randomly generated from the same model. The signal-to-noise-ratio, denoted by  SNR and defined as $\var(\beta^T(T)\bX)/\var(\epsilon)$, is also reported as it
is  an important  measure of the complexity of the varying coefficient  model associated with the tuning parameters $[t_1,t_2]$.

\begin{table}
\footnotesize
\caption{Average numbers of true positive (TP) and false positive (FP), and prediction error (PE) over $200$ repetitions and their robust standard deviations (in parentheses) for
the conditional-INIS, greedy-INIS and  fBIC approaches  under the varying coefficient model defined in  Example~\ref{exp1}.}\label{sim3.tab}
\begin{center}
\begin{tabular}{|c|c|ccc|ccc|ccc|} \hline
$[t_1,t_2]$ &SNR &\multicolumn{3}{|c}{Conditional-INIS} & \multicolumn{3}{|c}{Greedy-INIS}  &\multicolumn{3}{|c|}{fBIC} \\
& &TP & FP & PE &TP & FP & PE &TP & FP & PE \\ \hline
$[0,0]$ &$16.85$ & 4 & 0.54 &1.10  & 4 &  13.01   & 1.41 & 4 &0 &0.95   \\
        &        &(0)&(0.75)&(0.05)&(0)&  (3.73)  &(0.17) &(0) &(0) &(0.04)\\ \hline
$[2,0]$ &$3.66$ & 4 & 0.20 &0.78  & 4 &  0.41   & 1.10 & 4 &0.01 &1.12   \\
        &        &(0)&(0)&(0.06)&(0)&  (0)  &(0.05) &(0) &(0) &(0.05)\\ \hline
$[3,0]$ &$3.32$ & 4 & 0.19 &1.03  & 3.99 &  0.57   & 1.22 & 4 &0.01 &1.20   \\
        &        &(0)&(0)&(0.06)&(0)&  (0)  &(0.07) &(0) &(0) &(0.04)\\ \hline
$[2,1]$ &$3.21$ & 3.97 & 0.26 &1.27  & 3.90 &  1.14   & 1.63 & 4 &0 &1.20   \\
        &        &(0)&(0)&(0.24)&(0)&  (0)  &(0.41) &(0) &(0) &(0.07)\\ \hline
$[3,1]$ &$2.81$ & 3.95 & 0.31 &1.30  & 3.77 &  0.27   & 1.29 & 3.99 &0 &1.18   \\
        &        &(0)&(0.75)&(0.12)&(0)&  (0)  &(0.17) &(0) &(0) &(0.07)\\ \hline
\end{tabular}
\end{center}
\end{table}

 Table~\ref{sim3.tab}  displays the simulation results under the varying coefficient model defined in Example~\ref{exp1}.
 We can see that the fBIC in general outperforms both the conditional-INIS and the greedy-INIS approaches 
 in terms of the values of TP, FP, and PE.  In the first three cases where  $X_j$'s and $T$ are uncorrelated, all the three procedures are comparable in terms of selecting correctly all of the true covariates, but the fBIC selects fewer false covariates than the other two competitors and the fBIC also has smaller values of PE in general. In the latter two cases where $X_j$'s and $T$ are correlated, the performance of  the conditional-INIS and greedy-INIS approaches become worse while the performance of fBIC is still good in terms of the values of TP, FP, and PE.  The good performance of fBIC is  consistent with what we observed from Figure~\ref{BICvsEBIC1}.

\begin{table}
\footnotesize
\caption{The same as that of Table~\ref{sim3.tab} but now  under the varying coefficient model defined in  Example~\ref{exp2}.}\label{sim4.tab}
\begin{center}
\begin{tabular}{|c|c|ccc|ccc|ccc|} \hline
$[t_1,t_2]$ &SNR &\multicolumn{3}{|c}{Conditional-INIS} & \multicolumn{3}{|c}{Greedy-INIS}  &\multicolumn{3}{|c|}{fBIC} \\
& &TP & FP & PE &TP & FP & PE &TP & FP & PE \\ \hline
$[0,0]$ &$47.68$ & 8 & 0.21 &1.24  & 8 &  10.71   & 1.57 & 8 &0.02 &1.22   \\
        &        &(0)&(0)&(0.09)&(0)&  (3.73)  &(0.20) &(0) &(0) &(0.09)\\ \hline
$[2,0]$ &$9.40$ & 8 & 0.13 &1.17  & 8 &  0.60   & 1.16 & 8 &0 &1.20   \\
        &        &(0)&(0)&(0.09)&(0)&  (0)  &(0.10) &(0) &(0) &(0.08)\\ \hline
$[3,0]$ &$8.18$ & 7.90 & 0.10 &1.21  & 7.98 &  0.71   & 1.29 & 7.99 &0.03 &1.18   \\
        &        &(0)&(0)&(0.12)&(0)&  (0)  &(0.10) &(0) &(0) &(0.11)\\ \hline
$[2,1]$ &$8.62$ & 7.80 & 0.20 &2.16  & 7.55 &  0.26   & 2.26 & 8 &0.01 &2.55   \\
        &        &(0)&(0)&(0.58)&(0.75)&  (0)  &(0.70) &(0) &(0) &(0.64)\\ \hline
$[3,1]$ &$7.61$ & 7.75 & 0.18 &1.65  & 7.35 &  0.28   & 1.84 & 7.96 &0.02 &1.37   \\
        &        &(0)&(0)&(0.26)&(0.75)&  (0)  &(0.42) &(0) &(0) &(0.22)\\ \hline
\end{tabular}
\end{center}
\end{table}

Table~\ref{sim4.tab}  displays the simulation results under the varying coefficient model defined in Example~\ref{exp2}. Similarly,
 it is seen that fBIC in general outperforms the conditional-INIS and greedy-INIS approaches. Along with increases in the correlations between $X_j$'s and the correlations between $X_j$'s and $T$, the performance of the conditional-INIS and greedy-INIS  approaches become worse very quickly while the performance of fBIC becomes worse much more slowly.
 The good performance of the fBIC is  consistent with what we observed from Figure~\ref{BICvsEBIC2}.

\subsection{Applications to the Boston housing data}

Following \cite{FMD2014},  we  applied the fBIC approach to the well-known  Boston
housing dataset (Harrison and Rubinfeld 1978) whose description can be found in the manual of R
package {\it mlbench}. The dataset contains  $506$ census tracts
of Boston from the $1970$ census with $13$ covariates. The housing value equation obtained in the literature,
as reported by \cite{HR1978}, can be written as
\begin{equation}\label{boston1.equ}
\begin{array}{rcl}
\log(MV) &=& \beta_0 + \beta_1 RM^2 + \beta_2 AGE + \beta_3 \log(DIS) \\
&& + \beta_4\log(RAD)+ \beta_5 TAX +\beta_6 PTRATIO\\
&& + \beta_7 (B-0.63)^2  + \beta_8 \log(LSTAT) + \beta_9 CRIM\\
&& +\beta_{10} ZN + \beta_{11} INDUS + \beta_{12} CHAS\\
&& + \beta_{13} NOX^2+\epsilon,\\
\end{array}
\end{equation}
where the dependent variable $MV$ is the median value of owner-occupied homes, and the independent covariates
are quantified measurements of its neighborhood.  To adopt a varying coefficient model for the Boston housing data, \cite{FMD2014}  took  the covariate $\log(DIS)$, the weighted distance
to five employment centers in the Boston region, as the index variable $T$ and replaced the constant coefficients $\beta_j$  in (\ref{boston1.equ})
 with the varying coefficients $\beta_j(T)$. This allows us to examine
how the weighted distance to the five employment centers interacts with the other covariates. It seems reasonable to assume that the
impacts of the other covariates on housing price change with this distance. Using the conditional-INIS approach, \cite{FMD2014} obtained
the following varying coefficient submodel:
\begin{equation}\label{boston2.equ}
\begin{array}{rcl}
\log(MV) &=& \beta_0(T) + \beta_1(T) RM^2 + \beta_2(T) AGE + \beta_5(T) TAX\\
&& +\beta_7(T) (B-0.63)^2 + \beta_9(T) CRIM+\epsilon.
\end{array}
\end{equation}
By the fBIC  approach, we obtained  the following varying coefficient submodel:
\begin{equation}\label{boston3.equ}
\begin{array}{rcl}
\log(MV) &=& \beta_0(T) + \beta_1(T) RM^2 + \beta_2(T) AGE \\
&& + \beta_6(T) PTRATIO + \beta_7(T) (B-0.63)^2\\
&& + \beta_8(T) \log(LSTAT) + \beta_9(T) CRIM\\
&& +\beta_{13}(T) NOX^2+\epsilon.
\end{array}
\end{equation}

It is interesting to compare the two varying coefficient submodels (\ref{boston2.equ}) and   (\ref{boston3.equ}) selected by the conditional-INIS approach of \cite{FMD2014} and the fBIC procedure respectively. We can see that model (\ref{boston3.equ})  does not introduce the covariate $TAX$ which is introduced in model (\ref{boston2.equ}),  while it includes three other covariates $PTRATIO, \log(LSTAT)$, and $NOX^2$ which are not present in model (\ref{boston2.equ}).  Notice that the covariate $PTRATIO$ denotes the pupil-teacher ratio by the town school district, and a lower ratio indicates each student receives more individual attention. It is reasonable that parents usually want to buy houses near good schools which tend to have smaller values of $PTRATIO$. Therefore, it is expected that $PTRATIO$ should have important negative impact on housing values. Notice also that  the covariate $LSTAT$ is the proportion of the population that is of lower status. It is natural that a larger proportion of poor people in a region often means lower average housing prices in that region. Therefore, $LSTAT$ should have important negative impact on the housing values. Finally notice that the covariate  $NOX$ is  a measure for air pollution
level, and it generally has a negative impact on the housing values since people usually want to live in a region where  there is less air pollution.  In summary, introduction of these three covariates in the model (\ref{boston3.equ}) sounds reasonable. In fact the correlations between the covariates $PTRATIO, \log(LSTAT)$, and $NOX^2$ and the response $\log(MV)$ are $-0.5017, -0.8230$, and $-0.4965$ respectively.
As for the covariate $TAX$, there is no doubt that it is an important covariate which may have important negative  impact on the housing evaluation; in fact, the correlation between $TAX$ and $\log(MV)$ is $-0.5615$. On the other hand, it also has strong
correlations with $PTRATIO, \log(LSTAT)$, and $NOX^2$, which are $0.5224,0.4609$, and $0.6415$ respectively. Therefore, with introduction of $PTRATIO, \log(LSTAT)$, and $NOX^2$ in the model already, the effect of $TAX$ on $\log(MV)$ may have been represented by that of $PTRATIO, \log(LSTAT)$, and $NOX^2$.

\begin{figure}[ptbh]
\centerline{\includegraphics[scale=0.50]{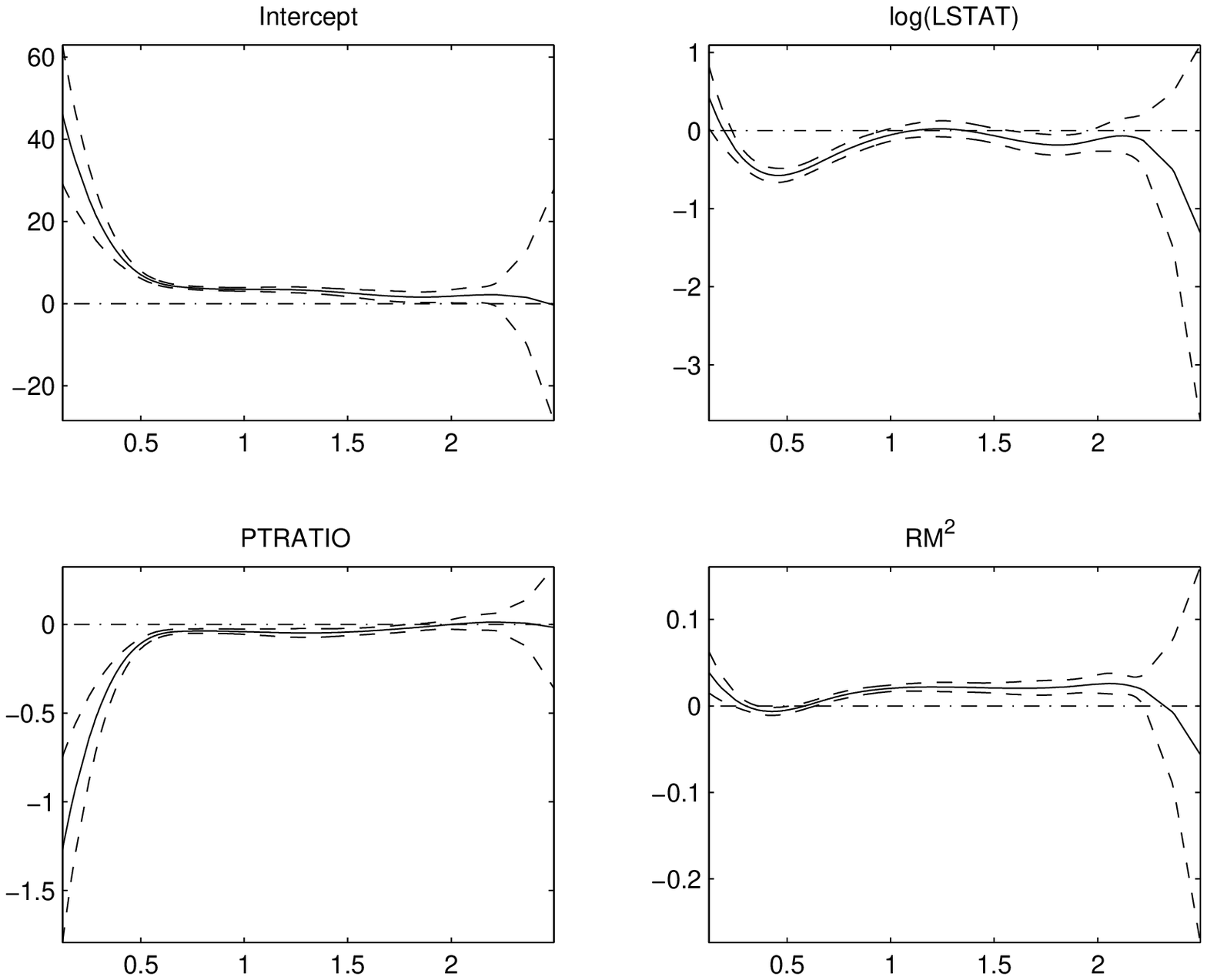}}
\centerline{\includegraphics[scale=0.50]{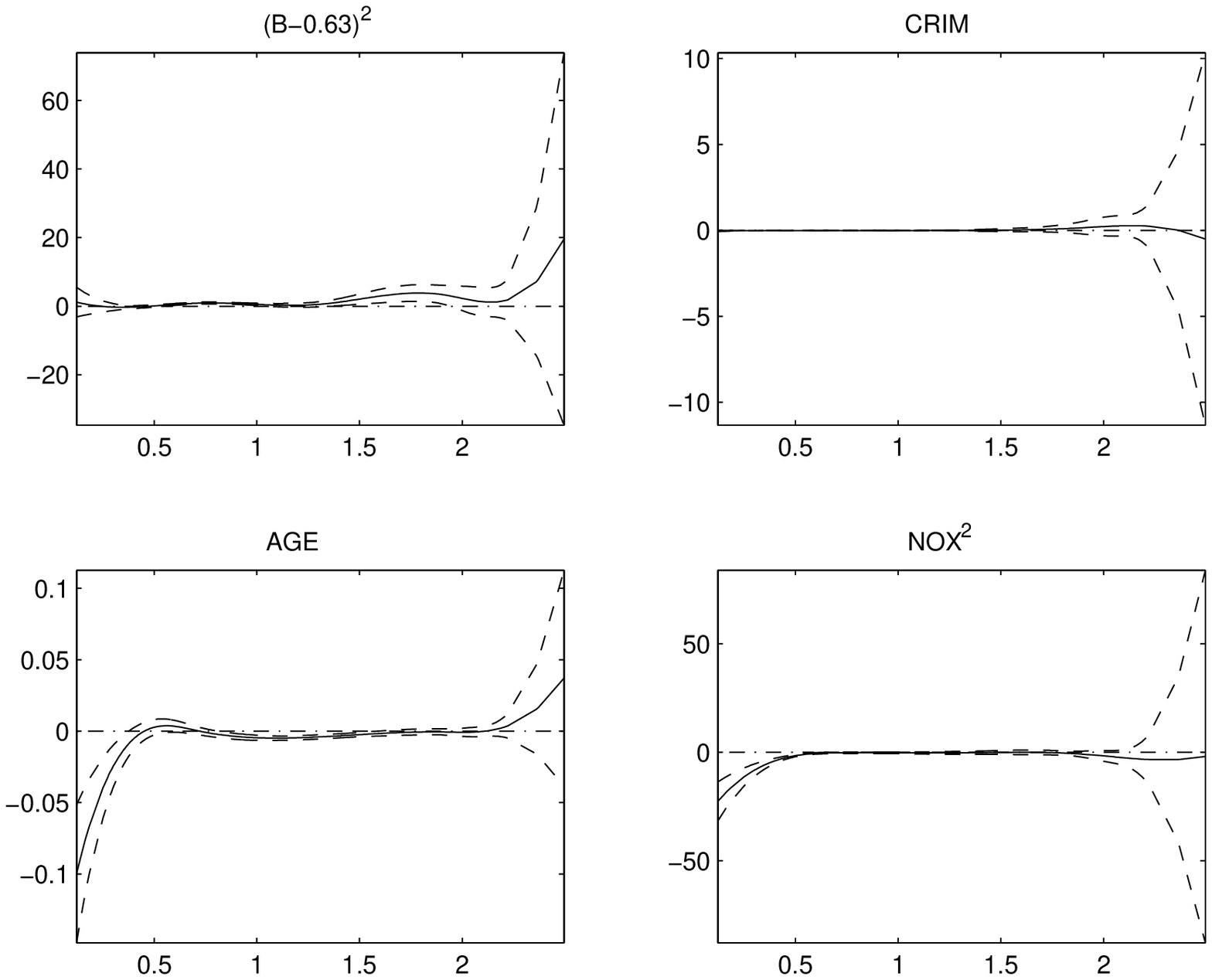}}
\caption{\em Fitted coefficient functions (solid) with approximate 95\% confidence bands (dashed) for  the Boston housing data. Cubic B-splines with the number of basis functions, $L_n=7$,  selected by fBIC, were used.} \label{housing.fit1}
\end{figure}

\begin{figure}[ptbh]
\centerline{\includegraphics[scale=0.35]{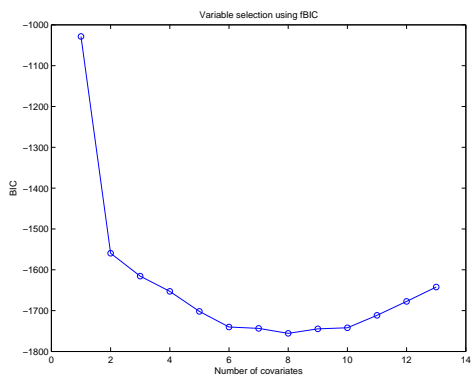}}
\caption{\em Variable  selection using fBIC for the Boston housing data.} \label{housing.fit2}
\end{figure}

 Figure~\ref{housing.fit1} plots the fitted coefficient functions $\beta_j(T)$'s, along with the corresponding approximate 95\%  confidence bands, according to the order in which they were selected by the fBIC, that is, the covariate $\log(LSTAT)$ was first selected, followed by the covariate $PTRATIO$, and then $RM^2$, etc.  Figure~\ref{housing.fit2} displays the BIC curve for the forward variable selection when applied to the Boston housing data.
 From Figure~\ref{housing.fit1}, it is seen that  the introduction of  $\log(LSTAT)$ in the model (\ref{boston3.equ}) at the first selection step indicates
that it has the most important impact on the housing values in the Boston regions under consideration, and
the socioeconomic status  distinctions mean  more in the upper brackets of the society than in the lower classes. The associated  coefficient curve shows that the impact of $\log(LSTAT)$ on housing values is generally negative as expected, especially when the regions are near the five employment centers. The effect at both ends are not significant and may be due to boundary effect of B-spline smoothing when less data are available.  The  introduction of $PTRATIO$ at the second step indicates that this covariate also has important impact on the housing value.  The associated coefficient curve shows that  the impact is negative, especially at those regions near the five employment centers.  The covariate  $RM$  is the third covariate introduced in the model (\ref{boston3.equ}), and it is the average number of rooms in owner units, which represents the size of a house. As expected, this covariate has positive impact on the housing value.  The impacts of the other four selected covariates on housing values can be analyzed and interpreted similarly; see \cite{FMD2014} and \cite{HR1978} for more details.

The Boston housing data set has only twelve covariates under consideration with $\log(DIS)$ as the index covariate. It  can not be regarded as  a real high-dimensional  data example.
To overcome this difficulty, \cite{FMD2014} extended the Boston housing data via introducing the
following artificial covariates:
\[
X_j =\frac{Z_j + 2U}{3}, j =13, 14,\cdots, 1000,
\]
where $Z_j, j=13,\cdots, 1000\iidsim N(0,1)$ and $U\sim U[0,1]$ are independent. They
randomly selected $n = 406$ observations  as the training set and applied their conditional-INIS and greedy-INIS approaches to select the models, and then computed the associated prediction mean squared error (PE)
on the rest $100$ observations. This process was  repeated $N = 100$ times and they reported the
average prediction error and model size, and their robust standard deviations as in Table~\ref{housing1.tab}. We repeated the above process with the fBIC approach and the results are also displayed in the table.
It turns out that the fBIC approach selects a few  artificial covariates. This is consistent with those observed in Figures~\ref{BICvsEBIC1} and \ref{BICvsEBIC2}.

To overcome this difficulty, we can first rank the covariates according to the BIC values of their corresponding marginal models
, and  then apply the fBIC approach to the  data with the first fifty covariates, say. 
The associated approach is called the modified fBIC approach. Since the dimensionality becomes smaller and it is expected that the fBIC approach will perform better in this case. The results presented in Table~\ref{housing1.tab} indicate that the average model size selected by the modified fBIC approach is indeed better than that selected by the fBIC approach, and it is about the same as that of model (\ref{boston3.equ}) which is selected when there are only twelve covariates involved. In addition, the PE and SNV values show that the modified fBIC approach improves on the fBIC approach substantially and that it is comparable with the Conditional-INIS and the Greedy-INIS.   Alternatively, as mentioned in Section \ref{screening}, we may apply the fBIC approach first and then apply the group SCAD to further remove those unwanted covariates. The resulting approach may be termed as the fBIC-SCAD approach, and the associated simulation results are listed at the last row of Table~\ref{housing1.tab}. The results show that applying group SCAD indeed improves the performance of the fBIC approach.

\begin{table}
\footnotesize
\caption{Prediction error (PE), model size (MS), and selected noise variables (SNV) over $100$ repetitions and their robust standard deviations (in parentheses) for the conditional-INIS, greedy-INIS, fBIC, modified fBIC approaches.}\label{housing1.tab}
\begin{center}
\begin{tabular}{|l|c|c|c|} \hline
 Approach &PE     & MS & SNV  \\ \hline
 Conditional-INIS & 0.046 (0.048)  & 5.55 (0.75)  & 0 (0)\\
 Greedy-INIS      & 0.048 (0.020)  & 4.80 (1.49)  & 0.01 (0)\\
 fBIC             & 0.083 (0.033)  & 8.60 (2.24)  & 2.16 (1.49)\\
 Modified fBIC    & 0.049 (0.019)  & 7.28 (1.49)  & 0.63 (0.75)\\
 fBIC-SCAD   & 0.062 (0.023)  & 7.00 (1.49)  & 1.89 (1.49)\\ \hline
\end{tabular}
\end{center}
\end{table}

From this example,
it is seen that  the fBIC approach  or  its modified version is  very useful in scientific discoveries based on high-dimensional data with complex structure. It  can select a
parsimonious close-to-truth model, and can reveal interesting relationship between the response variable and the important covariates.

\section{Proofs}
\label{sec:proofs}


First, we define some notation
related to the approximate regression models (\ref{eqn:e205}) and
(\ref{eqn:e207}). 
Let
\begin{align*}
D_{lSn}
& = n^{-1}\bm{W}_{S(l)}^T \bm{W}_{S(l)}
\quad \mbox{and}\quad D_{lS} = \RE \{ D_{lS_n} \},\\
d_{lS_n}
& = n^{-1}\bm{W}_{S(l)}^T \bm{Y}
\quad \mbox{and}\quad d_{lS} = \RE \{ d_{lS_n} \}, \quad \mbox{and} \\
\Delta_{lSn}
& = D_{lSn}^{-1}d_{lSn}- D_{lS}^{-1}d_{lS}\,.
\end{align*}
Then, the parameter vector $\overline\bgamma_l$ in model (\ref{eqn:e207}) can be expressed as $\overline \bgamma_l = (\bm{0}_L,\ldots,\bm{0}_L, \bI_L) D_{lS}^{-1}d_{lS}$, where $\bm{0}_L$ denotes the $L\times L$ zero matrix and $\bI_L$
is the $L$-dimensional identity matrix.

Before we prove Theorems \ref{thm:thm1} and \ref{thm:thm2}, we present
Lemmas \ref{lem:lem1}-\ref{lem:lem3}.  We verify these lemmas at the end of this section. In Lemma \ref{lem:lem1} we evaluate the minimum and maximum eigenvalues of some matrices.

\begin{lem}
\label{lem:lem1}
Assume that Assumptions {\bf T}, {\bf X}, and E(1) hold. Then, with
probability tending to 1,
there are positive constants $M_{11}$, $M_{12}$, $M_{13}$,
and $M_{14}$ such that
\begin{align*}
L^{-1}M_{11}
& \le \lambda_{\rm min} (D_{lSn}) \le \lambda_{\rm max}
(D_{lSn}) \le L^{-1}M_{12}\\
\intertext{and}
L^{-1}M_{13}
& \le \lambda_{\rm min} ( n^{-1}\bm{\widetilde W}_{lS}^T
\bm{\widetilde W}_{lS} ) \le \lambda_{\rm max}( n^{-1}
\bm{\widetilde W}_{lS}^T \bm{\widetilde W}_{lS})
\le L^{-1}M_{14}
\end{align*}
uniformly in $ S \subsetneq S_0$ and $l\in S^c$.
\end{lem}

Lemma \ref{lem:lem2} is about the relationship between
$\beta_l$ and $ \overline \bgamma_l$
in the extended marginal models (\ref{eqn:e201}) and (\ref{eqn:e205}).

\begin{lem}
\label{lem:lem2}
Assume that Assumptions {\bf T}, {\bf X}, and B(4)-(5) hold. Then
there are positive constants $M_{21}$ and $M_{22}$
such that
\[
M_{21}\sqrt{L}\big(\| \overline \beta_l \|_{L_2} - O(L^{-2})\big)
\le | \overline \bgamma_l | \le
M_{22}\sqrt{L}\big(\| \overline \beta_l \|_{L_2} + O(L^{-2})\big)
\]
uniformly in $ S \subsetneq S_0$ and $l\in S^c$.
\end{lem}

We use Lemma \ref{lem:lem3} to evaluate the estimation error for $\overline\bgamma_j$, $j\in S(l)$, in model (\ref{eqn:e205}).

\begin{lem}
\label{lem:lem3}
Assume that Assumptions {\bf T}, {\bf X}, and B(4)-(5) hold. Then, for any $\delta>0$,
there are positive constants $M_{31}$, $M_{32}$, $M_{33}$,
and $M_{34}$ such that
\[ | \Delta_{lSn} | \le M_{31}L^{3/2}p_0^{3/2}\delta/n
\]
uniformly in $ S \subsetneq S_0$ and $l\in S^c$, with probability
\[
1-  M_{32}\,p_0^2 \,L \exp \Big\{
-\frac{\delta^2}{M_{33}nL^{-1}+ M_{34}\delta}
+\log p + p_0 \log 2 \Big\}\, .
\]
\end{lem}

\subsection{Proofs of Theorems \ref{thm:thm1} and \ref{thm:thm2}, and Proposition \ref{prop:prop1} }

Now we prove Theorems \ref{thm:thm1} and \ref{thm:thm2}
by employing Lemmas \ref{lem:lem1}-\ref{lem:lem3}.\\

\noindent {\bf Proof of Theorem \ref{thm:thm1}.}
Consider the case that $ S \subsetneq S_0$ and $l\in S^c$. Note we can write
\begin{equation}
\widehat \bgamma_l =  \overline\bgamma_l + ({\bm 0}_L, \ldots, {\bm 0}_L, \bI_L) \Delta_{lSn}.
\label{eqn:e501}
\end{equation}
Lemma \ref{lem:lem1} implies we should deal with $\Delta_{lSn}$ on the right-hand side
of (\ref{eqn:e501}) when we evaluate $\widehat\sigma^{2}_{S}-\widehat\sigma^{2}_{S(l)}$ given in equation (\ref{eqn:e251}). For this purpose, Assumption B(2) suggests that we should take $\delta$ in Lemma \ref{lem:lem3} as
$\delta = n^{1-c_\beta/4}\kappa_n/L$ tending to
$\infty$. Recall the definition
of $\kappa_n$ in Assumption B(2). Then we have
that
\begin{align}
& \frac{\sqrt{L}\kappa_n}{L^{3/2}p_0^{3/2} \delta / n }=
\frac{ n^{ c_\beta / 4 } }{ p_0^{3/2} } \to \infty
\label{eqn:e503}\\
\intertext{and}
\lefteqn{\displaystyle
p_0^2L \exp\Big\{
-\frac{1}{2M_{33}} \frac{\delta^2}{nL^{-1}}
+ \log p + p_0 \log 2  \Big\}}\label{eqn:e505}\\
 & =  p_0^2L \exp\Big\{
-(2M_{33})^{-1} n^{1-c_\beta/2} \kappa_n^2 L^{-1}
+ \log p + p_0 \log 2  \Big\}\nonumber \\
 & <  p_0^2L \exp\Big\{
-(2M_{33})^{-1} n^{c_\beta/2} \log p
+ \log p + p_0 \log 2  \Big\}\to 0.\nonumber
\end{align}

By (\ref{eqn:e503}), (\ref{eqn:e505}), and Lemma \ref{lem:lem3},
$({\bm 0}_L, \ldots, {\bm 0}_L, \bI_L) \Delta_{lSn}$ is negligible compared to
$\bgamma_l$ on the right-hand side of (\ref{eqn:e501}), with probability tending to 1.
Therefore Lemmas \ref{lem:lem1} and \ref{lem:lem2}
and Assumption B(3) imply that we should focus on
$\sqrt{L}\| \beta_l \|$ in evaluating
$\widehat \sigma_{S(l)}^2$ in (\ref{eqn:e251}).
Hence the desired result follows from Assumption B(1).
\hfill $\Box$
\vspace{0.1in}

\noindent {\bf Proof of Theorem \ref{thm:thm2}.}
\noindent
To prove result (i), we evaluate
\[
\mbox{EBIC}(S) - \mbox{EBIC}(S(l))
=n\log \Big(
\frac{n \widehat \sigma_S^2}{n \widehat \sigma_{S(l)}^2}\Big)
- L(\log n + 2\eta \log p).
\]
Since
\[
n \widehat \sigma_S^2 - n  \widehat \sigma_{S(l)}^2
= ( \bm{\widetilde W}_{lS}^T \bm{\widetilde Y}_{S})^T
( \bm{\widetilde W}_{lS}^T \bm{\widetilde W}_{lS} )^{-1}
( \bm{\widetilde W}_{lS}^T \bm{\widetilde Y}_{S})
= \widehat \bgamma_l^T
\bm{\widetilde W}_{lS}^T \bm{\widetilde W}_{lS}
\widehat \bgamma_l,
\]
we have
\begin{equation}
\frac{n \widehat \sigma_S^2}{n \widehat \sigma_{S(l)}^2}
\ge 1 +( n^{-1} \bm{Y}^T\bm{Y})^{-1}
\widehat \bgamma_l^T \Big(\frac{1}{n}
\bm{\widetilde W}_{lS}^T \bm{\widetilde W}_{lS}\Big)
\widehat \bgamma_l.
\label{eqn:e507}
\end{equation}
Then Lemma \ref{lem:lem1} and (\ref{eqn:e507}) imply that
we have for some positive $C$,
\begin{equation}
\mbox{EBIC}(S) - \mbox{EBIC}(S(l)) \ge
C nL^{-1}| \widehat \bgamma_l |^2 - L (\log n + 2 \eta \log p)
\label{eqn:e509}
\end{equation}
uniformly in $S \subsetneq S_0$ and $l\in S^c$,
with probability tending to 1. Here we use
the fact that
$ L^{-1} | \widehat \bgamma_l  |^2$
is uniformly bounded with probability tending to 1.
Then as in the proof of
Theorem \ref{thm:thm1}, we should consider
$\sqrt{L}\| \overline \beta_j \|$ in evaluating
the right-hand side of (\ref{eqn:e509}). Since Assumption B(2)
implies that
\[
\frac{nL^{-1}( \sqrt{L}\kappa_n )^2}{L(\log n + 2\eta \log p)}
= \frac{n\kappa_n^2}{L(\log n + 2\eta \log p)}\to \infty,
\]
we have from (\ref{eqn:e509}) that
\[
\mbox{EBIC}(S) - \mbox{EBIC}(S(l)) > 0
\]
uniformly in $S \subsetneq S_0$ and $l\in S^c$
satisfying
$\displaystyle{\| \overline \beta_l \|_{L_2} \big/ \max_{j \in S_0 - S}
\|  \beta_j \|_{L_2} > C_L}$, with probability
tending to 1. Hence the proof of result (i) is complete.

\noindent
To prove result (ii), recall that we replace Assumption E(1) with Assumption E(2).
We should evaluate
\begin{eqnarray}
\lefteqn{\mbox{EBIC}(S_0(l)) - \mbox{EBIC}(S_0)}\label{eqn:e511}\\
& = & n\log \Big\{1-
\frac{\bm{Y}^T \bm{\widetilde W}_{lS_0}
( \bm{\widetilde W}_{lS_0}^T \bm{\widetilde W}_{lS_0} )^{-1}
\bm{\widetilde W}_{lS_0}^T \bm{Y}
}{n \widehat \sigma_{S_0}^2}
\Big\}
+ L(\log n + 2\eta \log p)\nonumber
\end{eqnarray}
for $l\in S_0^c$. It is easy to prove
that $ \widehat \sigma_{S_0}^2 $ converges to
$\RE \{ \epsilon ^2 \}$ in probability and the details
are omitted.
We denote $\bm{\widetilde W}_{lS_0}
( \bm{\widetilde W}_{lS_0}^T \bm{\widetilde W}_{lS_0} )^{-1}
\bm{\widetilde W}_{lS_0}^T $ by $\widetilde \bP_{lS_0}$,
which is an orthogonal projection matrix. Thus, from (\ref{eqn:e511}) we have for
some positive $C$,
\begin{equation}
\mbox{EBIC}(S_0(l)) - \mbox{EBIC}(S_0) \ge
-\frac{C}{\RE \{ \epsilon ^2 \}}
\bm{Y}^T \widetilde \bP_{lS_0} \bm{Y}
+ L (\log n + 2 \eta \log p)
\label{eqn:e513}
\end{equation}
uniformly in $l\in S_0^c$, with probability tending to 1.

Now we evaluate $ \bm{Y}^T \widetilde \bP_{lS_0} \bm{Y} $ on the right-hand side of (\ref{eqn:e513}).
From the definition of $ \bm{\widetilde W}_{lS_0} $, we have
\[
\bm{Y}^T \widetilde \bP_{lS_0} \bm{Y} =
(\bm{Y} -  \bm{W}_{S_0}\bgamma_{S_0} )^T \widetilde \bP_{lS_0}
(\bm{Y} -  \bm{W}_{S_0}\bgamma_{S_0} )
\]
for any $\bgamma_{S_0} \in \RR^{L \# S_0}$. Therefore we obtain
\[
\bm{Y}^T \widetilde \bP_{lS_0} \bm{Y}
\le \bm{\epsilon}^T \widetilde \bP_{lS_0} \bm{\epsilon}
+ |\bm{b}|^2
\]
where $\bm{\epsilon} = (\epsilon_1, \ldots, \epsilon_n)^T$
and $\bm{b}$ is some $n$-dimensional vector of spline approximation
errors satisfying $| \bm{b} |^2 =O(nL^{-4})$ uniformly
in $l\in S_0^c$.
By applying Proposition 3 of \cite{Zhang2010}, 
we obtain
\begin{equation}
\RP\Big(
\frac{\bm{\epsilon}^T \widetilde \bP_{lS_0} \bm{\epsilon}}{LC_{E2}}
\ge \frac{1+x}{
\{1 - 2/(e^{x/2}\sqrt{1+x}-1)\}_+^2}\Big)\le
\exp (-Lx/2) (1+x)^{L/2},
\label{eqn:e514}
\end{equation}
where $\{ x\}_+ = \max \{0, x \}$. We take $x=\log (p^{2\eta}n)a_n/2$
with $a_n$ tending to $0$ sufficiently slowly. Then from the above inequality,
we have $ \bm{\epsilon}^T \widetilde \bP_{lS_0} \bm{\epsilon}
=o_p ( L\log (p^{2\eta} n) )$ uniformly in
$l\in S_0^c$.
Thus we have
\begin{equation}
\bm{Y}^T \widetilde \bP_{lS_0} \bm{Y}
=O(nL^{-4}) + o_p ( L\log (p^{2\eta} n) )
\label{eqn:e515}
\end{equation}
uniformly in $l\in S_0^c$. Hence the desired result follows
from (\ref{eqn:e513}), (\ref{eqn:e515}), and the assumption
that $L=c_Ln^{\kappa_L}$ with $\kappa_L \ge 1/5$. Note that, here we use the condition that $L\log n/\log p\to\infty$ when $\eta =0$, which is stated in Assumption B(2).
\hfill $\Box$
\vspace{0.1in}

\noindent {\bf Proof of Proposition \ref{prop:prop1}.}
The first result 
follows from
almost the same arguments as in the proof of Theorem \ref{thm:thm1}, thus we omit the proof. We just comment on proof of the second
one, which corresponds to result (ii) of Theorem \ref{thm:thm2}.
We should deal with $S$ such that $S_0 \subset S
\subset \overline S_0$ in the proof. Then we replace
$\widehat \sigma_{S_0}^2$ in (\ref{eqn:e511})
with $\widehat \sigma_{S}^2$ and replace
$\widetilde \bP_{lS_0}$ with $\widetilde \bP_{lS}$ everywhere. Nevertheless, we still have
$ \bm{\epsilon}^T \widetilde \bP_{lS} \bm{\epsilon}
=o_p ( L\log (p^{2\eta} n) )$ uniformly in $S$ and
$l\in S^c$ by exploiting (\ref{eqn:e514}).
There is no change about the B-spline approximation.
Thus we obtain the version of (\ref{eqn:e513}) and (\ref{eqn:e515})
with $S_0$ replaced by $S$, and the modified
(\ref{eqn:e513}) and (\ref{eqn:e515}) hold
uniformly in $S$. Hence the latter half of
Proposition \ref{prop:prop1} is established. Note that some
minor conformable changes to the assumptions are necessary.

\subsection{Proofs of lemmas}

We use the following inequalities in the proofs of Lemmas
\ref{lem:lem1}-\ref{lem:lem2}.
\begin{equation}
\frac{C_{S1}}{L} \le \lambda_{\rm min} ( \RE \{ \bB(T)\bB(T)^T \} )
\le \lambda_{\rm max} ( \RE \{ \bB(T)\bB(T)^T \} ) \le \frac{C_{S2}}{L},
\label{eqn:e517}
\end{equation}
where $C_{S1}$ and $C_{S2}$ are positive constants independent of $L$. See \cite{HWZ2004} for
the proof of (\ref{eqn:e517}).\\

\noindent {\bf Proof of Lemma \ref{lem:lem1}. }
Write
\begin{equation}
n^{-1}D_{lSn}= n^{-1}\sum_{i=1}^n
(\bm{X}_{iS(l)}\bm{X}_{iS(l)}^T) \otimes
(\bB(T_i)\bB(T_i)^T),
\label{eqn:e519}
\end{equation}
where $\bm{X}_{iS(l)}$ is the $i$th sample version of
$\bm{X}_{S(l)}$ and $\otimes$ is the kronecker product.
Note that (\ref{eqn:e517}), (\ref{eqn:e519}), and Assumption X(2)
imply that, for any $\delta>0$,
\begin{equation}
\frac{C_{1}}{L} \le \lambda_{\rm min} ( D_{lS} )
\le \lambda_{\rm max} ( D_{lS} ) \le \frac{C_{2}}{L}.
\label{eqn:e521}
\end{equation}
for some positive $C_1$ and $C_2$.
In addition, by exploiting the band-diagonal property of $D_{lSn}$
and $D_{lS}$ and an exponential inequality, we can
demonstrate that
\begin{equation}| D_{lSn} - D_{lS}| \le n^{-1}\delta p_0
\label{eqn:e522}
\end{equation}
uniformly in $S \subsetneq S_0$ and $l\in S^c$
with probability
\begin{equation}
1- C_3 p_0^2 L \exp\{ -\delta^2( C_4 nL^{-1}
+C_5\delta)^{-1} \} \times p\exp ( p_0 \log 2 ),
\label{eqn:e523}
\end{equation}
where $C_3$, $C_4$, and $C_5$ are positive constants independent of
$p_0$, $L$, $n$, $p$, and $\delta$.
When we take $\delta = n^{1-c_\beta /4}L^{-1}$,
the probability in (\ref{eqn:e523}) tends
to 0 and the former result follows since
$ \delta p_0/n = p_0n^{-c_\beta/4}/L= o(L^{-1}) $.
The latter result follows from the following
relationship between $D_{lSn}^{-1}$ and
$ n^{-1} \bm{\widetilde W}_{lS}^T \bm{\widetilde W}_{lS}$:
\[
D_{lSn}^{-1} =
\begin{pmatrix}
* & * \\
* & \big(n^{-1} \bm{\widetilde W}_{lS}^T \bm{\widetilde W}_{lS}\big)^{-1}
\end{pmatrix}
\,.
\]
\hfill $\Box$

\noindent {\bf Proof of Lemma \ref{lem:lem2}.}
Let $\{ b_j \}_{j\in S(l)}$ be a set of square integrable
functions on $[0,1]$. Then Assumption X(2) implies that
\begin{equation}
C_{X2}\sum_{j\in S(l)} \| b_j(T) \|^2
\le \| \sum_{j\in S(l)} X_j b_j(T) \|^2
\le C_{X3} \sum_{j\in S(l)} \| b_j(T) \|^2.
\label{eqn:e525}
\end{equation}
Besides, Assumption {\bf T} implies
\begin{equation}
C_{T1}\| b \|_{L_2}^2
\le \| b (T) \|^2
\le C_{T2} \| b \|_{L_2}^2
\label{eqn:e527}
\end{equation}
for any square integrable function $b$.
In addition, due to Assumptions B(4) and B(5),
we can choose some positive constant $C_1$ and a set of $L$-dimensional vectors
$\{ \tilde \bgamma_j \}_{j\in S(l)}$ such that
\begin{equation}
\sum_{j\in S(l)} \| \overline \beta_j -
\tilde \bgamma_j^T \bB\|_\infty \le C_1 L^{-2},
\label{eqn:e529}
\end{equation}
where $C_1$ depends only on the assumptions.

By exploiting (\ref{eqn:e525})-(\ref{eqn:e529}), we obtain
\begin{eqnarray*}
\lefteqn{C_{X2}\sum_{j\in S(l)} \| \overline \beta_j(T) -
\overline \bgamma_j^T \bB(T) \|^2 }\\
& \le & \| \sum_{j\in S(l)} ( \overline \beta_j(T)
-  \overline \bgamma_j^T \bB(T) )X_j \|^2
\, \le \, \| \sum_{j\in S(l)} ( \overline \beta_j(T)
-  \tilde \bgamma_j^T \bB(T) )X_j \|^2 \\
& \le & \sum_{j\in S(l)} \| \overline \beta_j(T) -
\tilde \bgamma_j^T \bB(T) \|^2
\, \le \,
C_{X3}\sum_{j\in S(l)} \| \overline \beta_j -
\tilde \bgamma_j^T \bB\|_\infty^2
\, \le \, C_{X3}C_1^2L^{-4}.
\end{eqnarray*}
Therefore, there is a positive constant $C_2$ such that
\[
\| \overline \beta_j(T) -
\overline \bgamma_j^T \bB(T) \| \le C_2L^{-2}.
\]
This implies that
\begin{equation}
\| \overline \beta_j(T) \| -C_2L^{-2}
\le \Big\{ \overline \bgamma_j^T
\RE \{ \bB(T)\bB(T)^T \} \overline \bgamma_j \Big\}^{1/2}
\le \| \overline \beta_j(T) \| +C_2L^{-2}.
\label{eqn:e531}
\end{equation}
The desired result follows from (\ref{eqn:e517})
and (\ref{eqn:e531}).
\hfill $\Box$
\vspace{0.1in}

\noindent {\bf Proof of Lemma \ref{lem:lem3}. }
Recall the notation defined at the beginning of this section.
First we deal with $|d_{lS}|$ and $|d_{lSn}-d_{lS}|$.

We have $|d_{lS}| \le C_1 (p_0/L)^{1/2}$ from the definition of
the B-spline basis. As in the proof of Lemma 2 of \cite{CHLP2014},
we have
\[ | d_{lSn} - d_{lS} | \le \delta (Lp_0)^{1/2}/n
\]
uniformly in $S \subsetneq S_0$ and $l\in S^c$
with probability
\[
1- C_2 p_0 L \exp\{ -\delta^2( C_3 nL^{-1}
+C_4\delta)^{-1} \} \times p\exp ( p_0 \log 2 ),
\]
where $C_2$, $C_3$, and $C_4$ are positive constants independent of
$p_0$, $L$, $n$, $p$, and $\delta$.

By combining the above results, (\ref{eqn:e522}),
and Lemma \ref{lem:lem1},
we obtain
\begin{align}|D_{lSn}^{-1}(d_{lSn}-d_{lS})|
&\le C_5 L^{3/2}p_0^{1/2} \delta /n \label{eqn:e533}\\
\intertext{and} | (D_{lSn}^{-1}- D_{lS}^{-1})d_{lS}|
&\le |D_{lS}^{-1}| | D_{lSn}
- D_{lS}| |D_{lSn}^{-1}| |d_{lS}|
\le C_5 L^{3/2}p_0^{3/2} \delta /n
\label{eqn:e535}
\end{align}
uniformly in $S \subsetneq S_0$ and $l\in S^c$,
with probability given in the lemma. Note that
$C_5$ is independent of $p_0$, $L$, $n$, and
$\delta$. Hence the desired result follows
from (\ref{eqn:e533}) and (\ref{eqn:e535}).
\hfill $\Box$


 
\end{document}